\documentclass[
preprintnumbers,amsmath,amssymb,
preprint,
aps, prb]{revtex4}
\usepackage{graphicx}

\begin{document}

\title{Spatial Imaging and Mechanical Control of Spin Coherence in Strained GaAs Epilayers}
\author{H. Knotz}
\author{A. W. Holleitner}
\author{J. Stephens}
\author{R. C.  Myers}
\author{D. D. Awschalom}
\affiliation{Center for Spintronics and Quantum Computation,
University of California, Santa Barbara, California 93106}
\date{\today}
\begin{abstract}
The effect of uniaxial tensile strain on spin coherence in n-type GaAs epilayers is probed using time-resolved Kerr rotation, photoluminescence, and optically-detected nuclear magnetic resonance spectroscopies.  The bandgap, electron spin lifetime, electron g-factor, and nuclear quadrupole splitting are simultaneously imaged over millimeter scale areas of the epilayers for continuously varying values of strain.  All-optical nuclear magnetic resonance techniques allow access to the strain induced nuclear quadrupolar resonance splitting in field regimes not easily addressable using conventional optically-detected nuclear magnetic resonance. 
\end{abstract}
\maketitle

Strain is an important parameter in crystals for band-structure engineering and has been extensively\cite{Aspnes:1977}$^{,}$\cite{Carlos:1986}$^{,}$\cite{Sopanen:1995} studied in III-V semiconductors. 
Recent work\cite{Kato:2004}$^{,}$\cite{Kato:2005}$^{, }$\cite{Crooker:2005}has shown that 
electron spins in GaAs and related compounds respond to strain dramatically.  In particular, the manipulation of the spin-orbit coupling in 
GaAs via strain may be used for the development of all-electrical spintronic devices\cite{Kato:2005}.  

Here we employ a mechanical 3-point bending vise to controllably and reproducibly tune the tensile strain from 0.0 - 0.2{\%} \textit{in-situ}, a typical range for strain engineered heterostructures. 
This geometry creates a continuous variation in the magnitude of the strain which can be spatially imaged.  We observe significant changes to the electron spin and charge dynamics in bulk n-type GaAs samples.  Additionally, we investigate the effect of strain on the nuclear quadrupolar 
resonances (NQR) of each nuclear species present in GaAs using all-optical nuclear magnetic resonance (NMR)\cite{Kikkawa:2000}$^{,}$\cite{Salis:2001}$^{,}$\cite{Salis:2002}. 

A series of samples\cite{samples:1} was grown using molecular beam 
epitaxy consisting of a semi-insulating (100) GaAs substrate, 100 nm undoped GaAs buffer 
layer, 400 nm Al$_{0.7 }$Ga$_{0.3}$As, and a final 500 nm n-type GaAs active layer.  The doping ranges from 2 $\times $ 10$^{16}$ to 1 $\times $ 10$^{18}$ cm$^{ - 
3}$, with the majority of data taken on samples doped between 4 and 6$\times 
$10$^{16}$ cm$^{ - 3}$ to 
maximize electron spin lifetimes\cite{Kikkawa:1998}.  A longer electron spin 
lifetime is advantageous for the all-optical NMR measurement\cite{Kikkawa:2000}.

The samples are held in a variable temperature magneto-optical cryostat and probed using time-resolved Kerr rotation (TRKR)\cite{Crooker:1995}$^{,}$ \cite{Kikkawa:2000}$^{,}$\cite{Salis:2001}$^{,}$\cite{Salis:2002}$^{,}$\cite{Kato:2004}$^{,}$\cite{Kato:2005}, 
photoluminescence (PL), and all-optical NMR spectroscopies.  All measurements are performed at T = 5 K unless otherwise specified.  Excitation is provided by 
a pulsed Ti:Sapphire laser with a repetition rate of 76 MHz and energy of 1.525 eV for the Kerr studies, 1.614 eV for the PL experiments.  The magnetic field is applied 
parallel to [011] (Figure 1).  Pump and probe pulses are focused on the sample with 
a spot diameter of $\sim $ 50 $\mu$m and powers of 1 mW and 100 $\mu$W, respectively.  The pump beam is modulated between left and right circular polarization with a photoelastic modulator (PEM) at 50 kHz to reduce\cite{This:1} the 
dynamic nuclear polarization (DNP)\cite{Lampel:1968}$^{,}$\cite{Paget:1977}, while the probe beam is linearly 
polarized and chopped at 1.16 kHz for lock-in 
detection.  Using a mechanical delay line, the delay time $\Delta t$ 
between the two pulses can be tuned from 0 to 6 ns.  The reflected probe polarization is rotated by an angle $\theta_{K}$, which is proportional to the electron spin polarization along the propagation direction.  $\theta_{K}$ as a function of delay time can be fit by  \textbf{$e^{\frac{-\Delta t}{T_{2}^{\ast }}}\cos{(\frac{\omega_{L}}{2 \pi}\Delta t)}$ }where $\Delta t$ is the pulse delay time, T$_{2}^{\ast }$ is the inhomogeneous transverse spin life-time and $\omega_{L}$ is the Larmor precession frequency of the electron spins. 

The sample is held against the upper 
supports with a weak adhesive layer at room temperature and cooled to T = 5 K 
without applying strain.  After thermal stability is achieved, strain is 
applied by forcing the wedge (Fig 1a) against the back of the sample.  This 
creates an approximately uniaxial strain along the x-direction that is compressive at the rear and tensile 
at the front of the sample with a neutral axis at the center.  Area scans are obtained by moving the objective 
lens using two piezo driven linear translators whose minimum step size is 0.2 $\mu$m.  Frequencies and lifetimes 
near the edge of the sample in the stressed state were consistent with 
measurements taken before straining, indicating that the edge restraints and 
adhesive were not introducing significant additional strain into the sample.  The onset of 
the strain can be seen as a change in the electron precession frequency, as 
in Fig 1(b), and can be reproducibly tuned to higher and lower levels.  For 
the majority of this work, after the initial straining, the sample is held 
below 30 K to avoid introducing additional strain.  All of the data 
presented in this paper are from the same sample in a 
single strained state but the effects were reproduced in a number of samples\cite{samples:1} across the abovementioned doping range.  

In Figure 1(b), typical TRKR data can be seen from the unstrained (open 
circle) and strained (solid line) areas of the sample.  Data from the 
region with highest strain show a significant change in both the spin lifetime and precession frequency.  There is an increase in the electron 
precession frequency due to the strain's effect on the electron 
g-factor\cite{Jagannath:1985}.  The magnitude of the change in the electron 
precession frequency in this case is 70 $\pm $ 5 MHz, corresponding to a 5{\%} shift towards more negative values of effective g-factor.  Suppression of the spin lifetime compared to the unstrained value is brought about by increased spin-splitting in the conduction band\cite{Safarov:1983}.

Spatially resolved PL reveals how the band gap changes over the sample area.  In Figure 2(b), spectra from several excitonic recombination peaks are visible.  
The blue (red) line is a double Lorentzian fit to the data from the unstrained
(strained) area of the sample.  The higher energy peak is recombinant luminescence from the neutral donor bound (D$^{0}$, X), ionized donor bound (D$^{ + }$, X), and free excitons 
(X)\cite{Ilegems:1985}$^{,}$\cite{Pavesi:1994} while the lower energy peak is 
composed of luminescence from recombination on carbon impurities, in particular 
the donor-acceptor transition (D$^{0}$, C$^{0}_{As}$) and conduction band to C 
acceptor (e, C$^{0}_{As}$)\cite{Ilegems:1985}$^{,}$\cite{Pavesi:1994}$^{ }$, amongst
others.  The application of tensile strain raises the valence band edge, visible 
as a red shift of the peaks to lower energy. Spatial variation of the energy 
shifts is shown in Fig 2(a).  Comparing the observed spectra with previous 
measurements of acceptor peak shifts versus strain,\cite{Kim:1997} we estimate 
the maximum strain to be $0.16{\%}$ in this sample. 

Figure 2(c) shows three line cuts from the images in Fig 2(a).  Symbols from 
each line cut are pictured in their respective images.  The chosen straining 
geometry allows for a variety of strain levels to be addressed without 
changing the physical setup.  Due to a small asymmetry in the sample jig, a 
two dimensional stress variation exists in the epilayer.  This variation can 
be seen in the spatially resolved maps of PL, precession frequency, and 
lifetime in Fig 2(a). 

All-optical NMR utilizes resonant depolarization of the nuclear 
spins that have been polarized via  
DNP to detect the 
electric quadrupolar resonances of the nuclei, which are sensitive to the 
local electric fields at atomic sites\cite{Carlos:1986}.  This technique is capable\cite{Poggio:2005} of addressing as few as 10$^{8}$ nuclear spins, and in contrast to detection of NQR with rf probes\cite{Buratto:1991}$^{-}$\cite{Kuhns:1997}, measurements can be performed within a large range of magnetic fields. 
In experiments using polarization resolved  
(PL)\cite{Guerrier:1994}$^{,}$\cite{Eickhoff:2003} to detect NQR, small 
magnetic fields are typically used ($\leq$ 1T) because the electron spin 
polarization (and thus the polarization detectable via PL), goes as 
\textbf{\textit{$\rho $}}\textbf{(B) $\propto$ }\textbf{1 
}\textbf{/(1+($\omega_{L}$T$_{2}^{\ast 
}$)$^{2})$}.\cite{Considered:1}  \textbf{$\omega $}$_{L}$ is nearly linear in 
magnetic field, thus the observed polarization  is diminished as 
the applied field increases.  Previous high field measurements\cite{Barrett:1994}$^{,}$\cite{Kuhns:1997} of NQR and traditional ODNMR\cite{Buratto:1991}$^{ , }$\cite{Guerrier:1994}$^{ , }$\cite{Eickhoff:2003} use an rf coil for excitation and/or detection of the NMR signal.  While these methods offer the possibility for pulse sequence manipulation, non-reliance on DNP, and wider temperature ranges, they are often complex to implement.

Measuring the Kerr rotation under continuous illumination as a function of 
lab time shows a change in electron precession frequency larger than that 
affected by the application of strain.  This is the result of the additional 
effective magnetic field created by the DNP which is significant despite the presence of the PEM in the optical path.  The DNP has a saturation time of 
45-60 minutes at these temperatures.  Figure 3(c) shows the evolution of the nuclear field as a 
function of lab time for three positions of increasing strain. 
The presence of DNP allows for the use of all-optical NMR to probe the 
behavior of the nuclei.  Of the three species present, $^{69}$Ga, $^{71}$Ga, 
and $^{75}$As, the $^{75}$As has the strongest resonance transitions.  This 
is due to both to $^{75}$As being the most abundant isotope, as well as the fact that 
the hyperfine coupling is strongest between the electron and the $^{75}$As 
atom\cite{Lampel:1968}.  For this reason, the majority of study was 
confined to that species.

NQR are detected\cite{Salis:2002} by measuring the Kerr rotation signal at a fixed delay of 1200 ps while sweeping the in-plane 
magnetic field.  As the applied field is swept through the 
resonant field for each species, its nuclear magnetization is destroyed by the alternating field of the pulsed laser. 
This causes a swift decrease in the total effective magnetic field felt by 
the electrons, and a correspondingly abrupt change in the Kerr rotation signal.  Fixing 
the pump-probe delay beyond 1 ns and at a point of high derivative of the Kerr 
rotation allows the precession frequency shift caused by the removal 
of the nuclear field to be more easily detected.  A sweep rate of 5 mT/minute provides time for the nuclear 
polarization to accumulate.  Figure 3(a) shows typical 
field scans from strained (solid line) and unstrained (open circle) areas of 
the sample.  B$_{0}$ (5.21T for $^{75}$As) is centered between the two 
transitions.  Two events\cite{Salis:2002} are visible in the data for each strain, corresponding to 
transitions from the m = $\pm $3/2 to m = $\mp$1/2 states.  A surprising 
observation is the significant change in field at which these transitions 
occur; their separation increases to 60 mT with the highest applied 
strain.  This corresponds to a splitting in nuclear precession frequency 
between the two quadrupolar transitions of $\sim $ 150 kHz.  By tuning the strain we are able to manipulate the nuclear precession frequency 
splitting between these two transitions in a reproducible and controllable 
fashion over a range of 10-150 kHz.  Figures 3(b) and 3(d) show the quadrupolar 
transition fields as a function of position along the X and Y directions, respectively.  The asymmetry in the sample vise can be seen clearly in the Y axis figure.

 The simplicity and sensitivity of the techniques used here present an attractive method for imaging electronic and nuclear maps of the strain fields in more complicated heterostructures.  The data show that strain-induced modification  of the spin-orbit coupling can occur on time scales commensurate with the electron spin lifetime.  This suggests that rapid mechanical modulation of strain in very high frequency structures\cite{Cleland:1996} may be able to actively control spin states in future high frequency spintronic devices.  

The authors would like to thank Y. K. Kato, M. Poggio, and J. Speck for their helpful 
insight and comments. This work was supported by the AFOSR, DARPA/DMEA, and the NSF.

\newpage
Figure captions

\begin{figure}[h]\caption{\label{fig1}
(a) Schematic of the experimental setup and stress geometry.  The laser is directed along the growth axis (z-axis).  A threaded rod (1) pushes the wedge (2) past the chip breaker (3) forcing it against the back of the sample.  Strain is produced (4) in the plane the sample, compressive at the rear and tensile in the active layer (black region).  (b) Kerr rotation as a function of delay time for the strained (solid line) and unstrained (open circle) areas of the sample (T = 5 K).  The Kerr rotation, $\theta_{K}$, is proportional to the conduction electron polarization along z.  $\omega_{L}$ can be obtained from the period of the reflected probe intensity, $\textbf{T}_{2}^{\ast}$ from the curve decay.  
}\end{figure}

\begin{figure}[h]\caption{\label{fig2}
(a) Top down, spatial maps of  C$^{0}_{As}$ exciton group PL, donor-bound exciton group PL,  effective g-factor, and $\textbf{T}_{2}^{\ast}$ over entire sample.  Symbols from line cuts in (c) pictured on images.  (b) PL from strained (square) and unstrained (circle) areas of the sample.  Overlaid are Lorentzian fits used to extract peak shift.  (c) Line cuts along X of C$^{0}_{As}$ group PL energy (solid square), effective g-factor (open square), and $\textbf{T}_{2}^{\ast}$ (open circle) from spatially resolved maps in Fig 2(a).  For the PL, the maximum shift at center is $\sim$ 10 meV.  The TRKR and PL measurements are performed at T = 5 K.
}\end{figure}

\begin{figure}[h]\caption{\label{fig3}
(a) NMR transitions for maximum (solid line) and minimum (open circle) strain. Arrows highlight resonant nuclear depolarization at (from left to right) $\Delta$B = -29 mT, -4 mT, +4 mT, and +29mT.  (b) X dependence of the NQR field  along the sample.  Pictured in center is a schematic of the line cut orientation.  Transitions are plotted in nuclear frequency shift (left) and magnetic field (right).  (c) Dynamically polarized nuclear field vs. strain as a function of laboratory time.  Square symbol is lowest strain, triangle is highest (T = 10 K).  (d) Y dependence of NQR field.  Unless noted, all measurements are performed at T = 5 K.
}\end{figure}

\newpage
\begin{center}

\vspace{1in}
Figure 1

Knotz \textit{et al.}
\vspace{1in}

\includegraphics{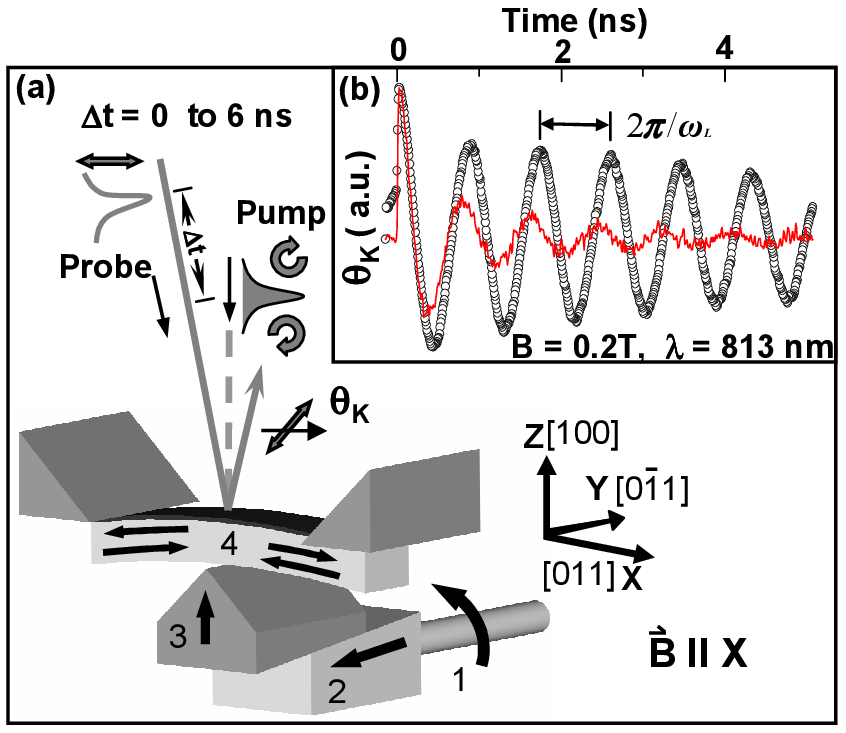}

\newpage

\vspace{1in}
Figure 2

Knotz \textit{et al.}
\vspace{1in}

\includegraphics{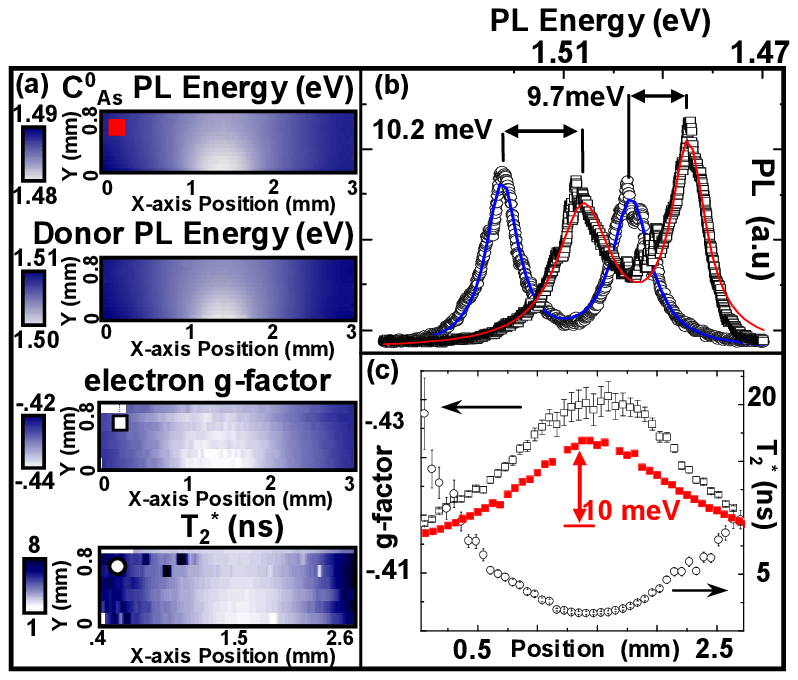}

\newpage

\vspace{1in}
Figure 3

Knotz \textit{et al.}
\vspace{1in}

\includegraphics{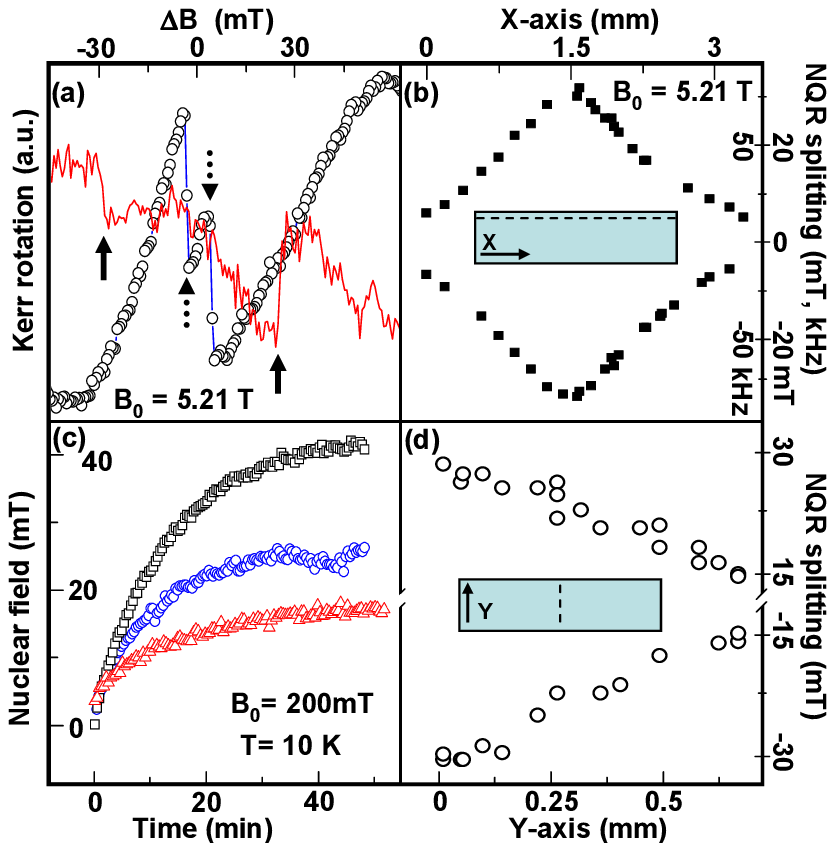}

\end{center}


\begin{thebibliography}{27}
\bibitem{samples:1} 14 samples in all were measured from 6 separate wafers.
\bibitem{This:1} This was desirable for the TRKR measurement, as the DNP creates a large additional effective magnetic field, greatly altering the observed precession frequency of the electrons.  The DNP could be removed by blocking the laser and reversing the applied field for a short time.
\end{thebibliography}
\end{document}